\documentclass[aps,prl,reprint,superscriptaddress]{revtex4-2}
\usepackage{graphicx}
\usepackage{xcolor}
\usepackage{dcolumn}
\usepackage{bm}
\usepackage[utf8]{inputenc}

\begin{document}
	
	\preprint{APS/123-QED}
	
	\title{Hysteretic Coherence Collapse Across the First Order CDW Transition in 1$T$-TaS$_2$}
	
	\author{Turgut Yilmaz}
	\email{trgt2112@gmail.com}
	\affiliation{Department of Physics, Xiamen University Malaysia, Sepang 43900, Malaysia}
	\affiliation{Department of Physics, University of Connecticut, Storrs, CT 06269, USA}

	\author{Anil Rajapitamahuni}
	\affiliation{National Synchrotron Light Source II, Brookhaven National Lab, Upton, New York 11973, USA}
	
	\author{Asish K. Kundu}
	\affiliation{National Synchrotron Light Source II, Brookhaven National Lab, Upton, New York 11973, USA}
	
	\author{Menka Jain}
    \affiliation{Department of Physics, University of Connecticut, Storrs, CT 06269, USA}
    \affiliation{Institute of Materials Science, Storrs, University of Connecticut, 06269, USA}
	
	\author{Elio Vescovo}
	\affiliation{National Synchrotron Light Source II, Brookhaven National Lab, Upton, New York 11973, USA}

	\date{\today}

\begin{abstract}

The first-order phase transition between the nearly commensurate (NC-CDW) and commensurate (C-CDW) charge-density wave phases in 1$T$-TaS$_2$ underpins its exotic electronic behavior, yet the spectroscopic evolution of the low-energy electronic structure across this transition remains crucial to understand. Using angle-resolved photoemission spectroscopy (ARPES), we investigate the low-temperature C-CDW phase, characterized by a flat band commonly associated with the lower Hubbard band and a distinct in-gap state located closer to the Fermi level. Photon-energy-dependent measurements distinguish these two low-energy features through their different spectral-weight evolution. Temperature-dependent ARPES across heating and cooling cycles reveals that the in-gap state undergoes an abrupt collapse upon heating into the NC-CDW phase and re-emerges sharply upon cooling back into the C-CDW phase. This pronounced thermal hysteresis provides direct spectroscopic evidence of the first-order nature of the transition. Furthermore, the disappearance and recovery of the in-gap state closely track the corresponding changes in resistivity, highlighting its intimate connection to the electronic reconstruction across the C-CDW–NC-CDW phase transition.

\end{abstract}

	\maketitle
	
	\section*{Introduction}
	
   The layered transition-metal dichalcogenide 1$T$-TaS$_2$ has long served as a paradigmatic platform for investigating the interplay between charge-density-wave (CDW) order, strong electronic correlations, and emergent quantum phases \cite{wilson1974charge, fazekas1979electrical}. This material exhibits a remarkable sequence of temperature-driven phase transitions: it transforms from an incommensurate (IC) CDW phase near 550~K to a nearly commensurate (NC) CDW phase around 350~K, and finally to a commensurate (C) CDW ground state below approximately 180~K upon cooling. \cite{thomson1994scanning, rossnagel2011origin}. The C-CDW phase is characterized by a distinctive $\sqrt{13} \times \sqrt{13}$ reconstruction, wherein 13 Ta atoms contract toward a central site to form Star-of-David clusters \cite{brouwer1980low}. This structural modulation is accompanied by the opening of an energy gap at the Fermi level, rendering the material insulating at low temperatures \cite{manzke1988electronic}.

	Furthermore, the CDW phase transitions in 1$T$-TaS$_2$ are accompanied by dramatic changes in electrical transport, most notably a pronounced, hysteretic metal-insulator like transition in resistivity between 180~K and 220~K upon cooling and heating, respectively \cite{di1977low,sipos2008mott}. This hysteretic behavior is a hallmark of the first-order C-CDW to NC-CDW transition and has attracted considerable attention for potential applications in memristive devices, electrical oscillators, and photodetectors \cite{yoshida2015memristive, stojchevska2014ultrafast, vaskivskyi2016fast,vaskivskyi2015controlling}.

    Despite decades of intensive study, the precise nature of the insulating ground state in 1$T$-TaS$_2$ remains a subject of vigorous debate. Early interpretations invoked a Mott–Hubbard mechanism, wherein electron correlations within the Star-of-David clusters localize the unpaired electron at the central Ta site, giving rise to lower and upper Hubbard bands \cite{fazekas1979electrical, pillo2000interplay}. Angle-resolved photoemission spectroscopy (ARPES) measurements consistently reveal a flat band at approximately 180~meV binding energy in the C-CDW phase, commonly attributed to the lower Hubbard band \cite{wang2020band,ritschel2015orbital}.
    
    However, in addition to this well-established feature, several studies have reported a distinct low-binding-energy electronic state located much closer to the Fermi level, residing within the nominal insulating gap~\cite{perfetti2005unexpected,jung2022surface,wang2024dualistic}. The existence of this in-gap state raises a fundamental question: if the C-CDW phase is a fully developed Mott insulator, what is the origin of this residual low-energy spectral weight? More recently, theoretical works have suggested that interlayer coupling and the specific stacking order of CDW layers may critically influence the electronic structure, potentially stabilizing band-insulating behavior under certain conditions \cite{ge2010first,darancet2014three,ritschel2018stacking,park2023stacking,lee2023charge}. Within this broader context, clarifying the nature and role of the low-energy in-gap state is essential for resolving the microscopic character of the insulating ground state.

\begin{figure*}[t]
	\centering
	\includegraphics[width=0.85\textwidth]{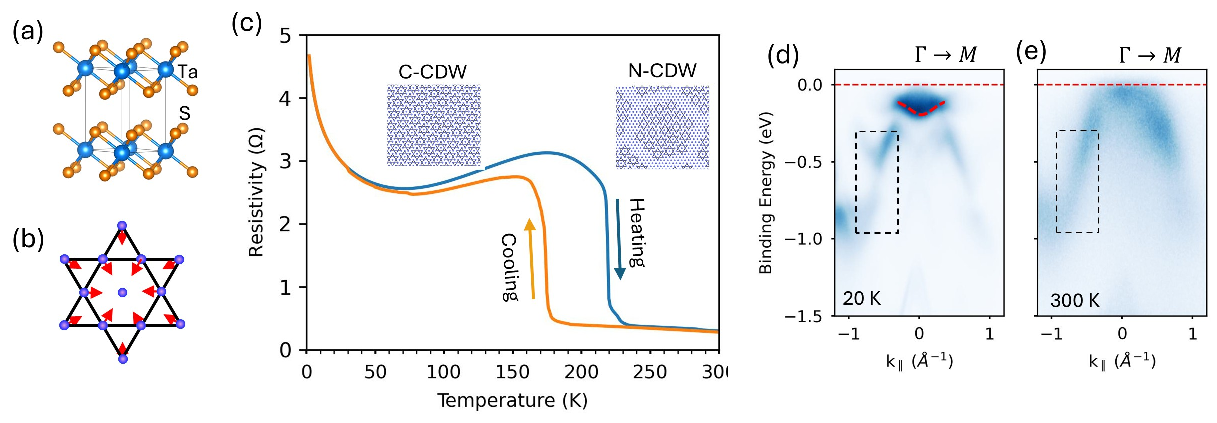} 
	
	\caption{\label{fig:epsart} 
		(a) Crystal structure of 1$T$-TaS$_2$ highlighting Ta (blue) and S (orange) atoms. 
		(b) Schematic of the star-of-David cluster illustrating atomic displacements associated with the commensurate charge density wave (C-CDW). 
		(c) Temperature-dependent resistivity measured upon cooling (orange) and heating (blue), showing hysteresis between the nearly commensurate (N-CDW) and commensurate (C-CDW) phases. Insets depict the corresponding atomic arrangements. 
		(d, e) ARPES intensity maps along $\Gamma$–M at 20 K (C-CDW phase) and 300 K (NC-CDW phase), respectively. The dashed line marks the Fermi level, and the red curve highlights the flat band in the C-CDW phase.}
	
\end{figure*}

     While the flat band at higher binding energy has been extensively characterized, the evolution of the in-gap state with temperature has received comparatively little attention. If this low-energy in-gap state persists within the nominal gap, it may contribute significantly to the transport and thus directly govern the hysteretic metal–insulator like transition. In this work, we employ photon-energy-dependent and temperature-dependent ARPES to investigate the electronic structure of 1$T$-TaS$_2$ across its C-CDW to NC-CDW phase transition. We study the temperature evolution of the in-gap electronic state located closer to the Fermi level than the well-known flat band. Most significantly, we track the evolution of this in-gap state through controlled heating and cooling cycles, revealing a clear hysteretic behavior that mirrors the resistivity hysteresis. The state disappears upon heating above $\sim$220~K and re-emerges upon cooling to $\sim$160~K, with its coherence peak near the Fermi level directly correlating with the insulating character of the low-temperature phase. Our findings establish this in-gap state as a defining low-energy electronic signature of the C-CDW phase in the phase transition, providing a unified explanation for the transport anomalies and opening new perspectives on the nature of the correlated ground state in 1$T$-TaS$_2$.

\begin{figure*}[t]
	\centering
	\includegraphics[width=0.85\textwidth]{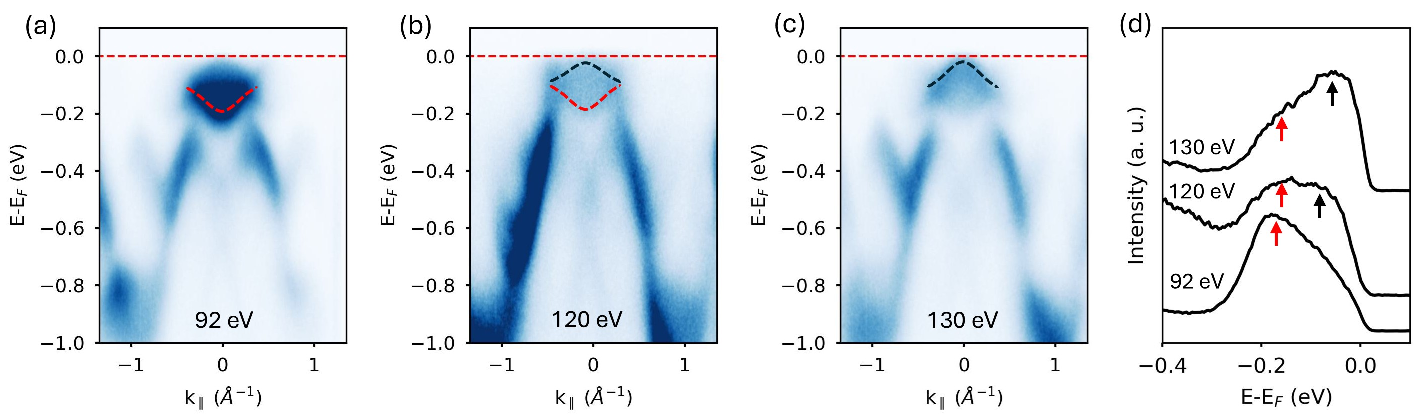} 
	
	\caption{\label{fig:epsart} 
		(a) Electronic band dispersion measured at 92 eV photon energy, with the dashed contour highlighting a flat band mear $-0.18$ eV. 
		(b) Band structure at 120 eV, where both red and black dashed contours emphasize the flat band and in-gap state, respectively. 
		(c) Band structure at 130 eV, with the black dashed contour marking the in-gap state. 
		(d) Corresponding energy distribution curves (EDCs) along the $\overline{\Gamma}$. Red and black arrows trace the flat band and in-gap state, respectively. ARPES data in panels (a)–(c) were recorded along the $\overline{\Gamma}$--$\overline{M}$ direction. Using the free-electron final-state model with an inner potential of $V_{0}=17.5$~eV~\cite{wang2020band}, the 92~eV and 120~eV measurements correspond approximately to the $\Gamma$ ($k_z=0$) and $A$ ($k_z=\pi/c$) planes, respectively.}

\end{figure*}

\begin{figure*}[t]
	\centering
	\includegraphics[width=0.85\textwidth]{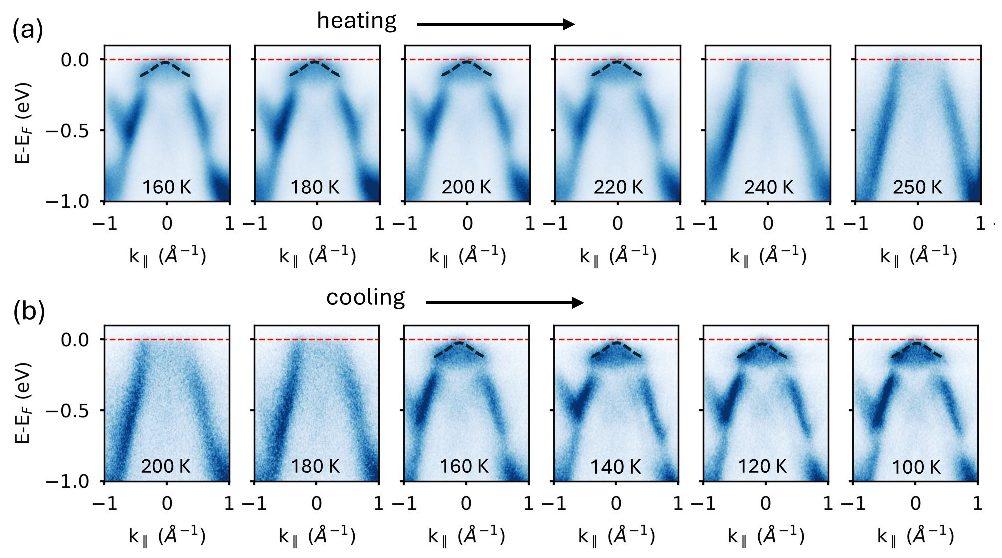} 
	
	\caption{\label{fig:epsart} 
		(a) ARPES intensity maps recorded upon heating from 160 K to 250 K, showing the progressive changes in the near-$E_F$ states. (b) Corresponding cooling sequence from 200 K down to 100 K, highlighting the hysteretic behavior of the band dispersion. The comparison of heating and cooling cycles highlights the path-dependent evolution of the electronic states across the transition. The data is collected with 130 eV photon energies along the $\overline{\Gamma}$--$\overline{M}$ high-symmetry direction. The dashed black curves highlight the in-gap state.}
	
\end{figure*}

\begin{figure*}[t]
	\centering
	\includegraphics[width=0.85\textwidth]{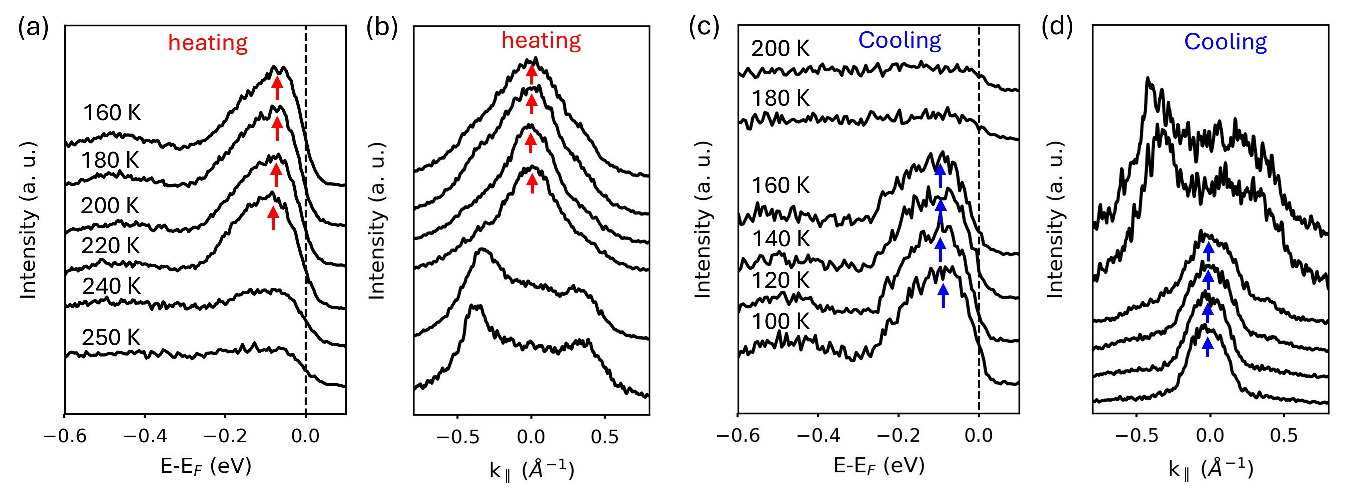} 
	
	\caption{\label{fig:epsart} 
		(a) Energy distribution curves (EDCs) recorded upon heating from 160 K to 250 K, showing the evolution of spectral weight near $E_F$. (b)  Momentum distribution curves (MDCs) along during heating, highlighting changes in band structure at the Fermi level. In (b), the same temperature scale is used as in (a). \textbf{a}. EDCs measured upon cooling from 200 K down to 100 K, revealing hysteretic behavior in the spectral features. (d)  MDCs along at the Fermi level during cooling. In (d), the same temperature scale is used as in (c). The data is taken from the ARPES maps given in Fig.~3. MDCs and EDCs were obtained by integrating the ARPES intensity within an energy window of 30 meV at the Fermi level and a momentum window of 0.05 \AA$^{-1}$ at k$_\parallel$ $=$ 0, respectively. Blue and red arrows mark the in-gap state.
	}
	
\end{figure*}

\section*{MATERIALS AND METHODS}

Single crystals of 1$T$-TaS\textsubscript{2} were obtained from 2Dsemiconductors. $\mu$-ARPES measurements were carried out at the ESM beamline (21-ID-1) of National Synchrotron Light Source II(NSLS-II) at at Brookhaven National Laboratory ~\cite{rajapitamahuni2024electron}, employing a DA30 Scienta electron spectrometer. The samples were cleaved \textit{in situ}, and the base pressure in the photoemission chambers was maintained below $3 \times 10^{-11}$~Torr. At 21-ID-1, the incident photon beam was focused to a spot size of approximately $5~\mu\mathrm{m}^2$, with synchrotron radiation impinging on the sample surface at an angle of $55^\circ$. Temperature dependent resistance measurements were performed using the dc resistivity option in Evercool Physical Property Measurement System (Quantum Design Inc.). The data was recorded in sweep mode while both heating and cooling the sample at the rate of 1 K/min.

\subsection{RESULTS}

    \subsection{Structural, Transport, and Electronic Properties Across the Charge-Density-Wave Phases}
	
	Fig.~1 summarizes the structural, transport, and electronic properties of the layered 1$T$-TaS$_2$ compound across its CDW phase transitions. The crystal structure, shown in Fig.~1a, consists of layered Ta atoms coordinated octahedrally by S atoms. The material forms two-dimensional sheets stacked along the out of plane direction with weak van der Waals interlayer coupling. In the low-temperature C-CDW phase, the Ta atoms undergo a periodic lattice distortion forming a characteristic Star-of-David superstructure, illustrated in Fig.~1b. In this reconstruction, 13 Ta atoms contract toward a central Ta site, producing a $\sqrt{13} \times \sqrt{13}$ superlattice relative to the undistorted lattice.

   	The temperature dependence of the electrical resistivity is shown in Fig.~1c for both cooling and heating cycles. The first transition during the heating happens around 220~K  from C-CDW to NC-CDW. Upon cooling, this transition is observed around 170~K. Therefore, the resistivity exhibits a clear thermal hysteresis.
	
     Fig.~1d, 1e present ARPES spectra acquired at 20~K and 300~K, corresponding to the C‑CDW and NC‑CDW phases, respectively. The measurements were taken along the $\Gamma \rightarrow M$ direction of the normal-phase Brillouin zone. At low temperature (20~K), within the C‑CDW phase, the spectra reveal a suppression of spectral weight near the Fermi level and a relatively flat feature at the $\Gamma$-point with a binding energy of $\sim$180~meV. Although this band is often attributed to the lower Hubbard band of a Mott-like phase, its precise origin remains elusive~\cite{ritschel2015orbital}. To remain consistent with the literature, we refer to it simply as the flat band. In addition, hybridization induced spectral weight suppression due to band folding is highlighted by the dashed rectangle in Fig.~1d. By contrast, in the data taken at 300~K and within the NC-CDW phase, the flat band is absent and the hybridization gaps are filled. The electronic states appear broader and more metallic, with enhanced spectral weight at the Fermi level. Furthermore, at 300 K, finite spectral weight remains at the $\Gamma$ point, originating from the underlying $\Gamma$ conduction band whose spectral weight is redistributed toward Fermi level by the NC-CDW reconstruction~\cite{yilmaz2026electronic}.

    \subsection{Photon-Energy-Dependent ARPES and Identification of an In-Gap State}
    
    Fig.~2 presents photon energy dependent ARPES measurements highlighting the presence of an in-gap electronic state. At 92~eV , the flat band is most prominently observed near the Brillouin zone center (Fig.~2a). Upon increasing the photon energy to 120~eV, the band structure resolves into two distinct states near the Fermi level (Fig.~2b). The higher binding energy feature corresponds to the flat band, while the state located closer to $E_F$ is the in-gap state marked with a dashed black line in Fig.~2b. At 130~eV (Fig.~2c), the in-gap state exhibits stronger spectral weight compared to the flat band. 

    The photon-energy dependence of the spectral weight for both states is further illustrated in Fig.~2d, which shows representative energy distribution curves (EDCs). The EDCs taken with 120 eV and 130 eV photon energies clearly resolve two peaks corresponding to the flat band at higher binding energy and the in-gap state closer to the Fermi level.
    
    The assignment of the flat band and the in-gap state is based on several independent experimental observations. First, the two features exhibit markedly different photon-energy dependences: the flat band is strongly enhanced at 92 eV, whereas the in-gap state reaches maximum intensity near 130 eV. Second, at intermediate photon energies (120 eV), both features are simultaneously resolved as two separate peaks in both the ARPES intensity maps and the corresponding EDCs, demonstrating that they cannot be described as a single electronic state with varying photoemission matrix elements. Third, the two features exhibit different temperature evolution across the first-order C-CDW–NC-CDW transition, with the in-gap state showing an abrupt hysteretic collapse while the flat band remains observable.

    \subsection{Temperature-Dependent ARPES Evolution Across Heating and Cooling Cycles}

     We now turn to the question of how the in-gap state evolves with temperature. Fig.~3 presents a systematic comparison of ARPES measurements acquired during controlled heating and cooling cycles. In the heating sequence (Fig.~3a), spectra acquired between 160~K and 250~K reveal a systematic evolution of the electronic structure near the Fermi level. As the temperature increases, the in-gap spectral feature does not gradually weaken but instead undergoes a sharp suppression, disappearing abruptly above 220~K. Simultaneously, the hybridization induced gaps in the band structure weakens, and the valence band evolves into a more continuous dispersion.

    In contrast, the cooling sequence (Fig.~3b), spanning temperatures from 200~K down to 100~K, demonstrates the re-emergence of the in-gap state at 160~K as the system returns to the low-temperature phase. Notably, the onset temperature for the reappearance of the in-gap feature does not coincide with the temperature at which it disappears during heating. This mismatch defines a clear hysteresis loop in the electronic structure, indicative of a first-order phase transition. Such hysteretic behavior is consistent with transport measurements, particularly resistivity as a function of temperature, which also exhibit looped characteristics across the transition. The strong correlation between the spectroscopic and transport signatures demonstrates that the temperature dependence of the in-gap state plays a crucial role in governing the transport properties of the material.

    \subsection{Temperature-Dependent EDC and Momentum Profiles During Heating and Cooling}

    The heating and cooling cycle analysis establishes that the in-gap state exhibits sharp hysteretic behavior, directly linking its presence to the underlying phase transition. To gain a more quantitative perspective on this evolution, we next examine the temperature dependence of EDCs and momentum distribution curves (MDCs). Figs~4(a,b) correspond to the heating sequence, while Figs~4c,d show the cooling sequence. EDCs are obtained along the $\overline{\Gamma}$-point. 
    
    The coherence peak of the in-gap state emerges in the vicinity of the Fermi level, contributing spectral weight directly at $E_F$. As the sample temperature increases, this coherent peak is abruptly suppressed and disappears above 220~K (Fig.~4a). This behavior is further confirmed by the momentum distribution curves, where the spectral bump accumulated around the $\overline{\Gamma}$-point at low temperatures transforms into two distinct peaks located symmetrically on opposite sides of the $\overline{\Gamma}$-point.

    Figs~4c, 4d present the cooling cycle from 200~K down to 100~K. The EDCs in Fig~4c demonstrate the reappearance and progressive strengthening of the in-gap state peak as the temperature decreases toward 160~K. At lower temperatures, the peak sharpens and gains spectral weight, signaling the recovery of the ordered phase. The corresponding momentum profiles in Fig~4d further confirm this behavior as the reverse of the heating process, the two distinct peaks observed near the Fermi level evolve into a single spectral bump around the $\overline{\Gamma}$-point at 160~K. The comparison between heating and cooling sequences thus highlights a clear hysteretic response in the electronic structure, consistent with transport measurements that exhibit similar looped characteristics across the transition.

\section{Discussion}

Temperature-dependent ARPES data reveal that the low-energy in-gap state, long considered secondary or extrinsic, is in fact a defining spectroscopic feature of the C-CDW phase. Its coherence peak collapses abruptly upon the transition to the NC-CDW phase and re-emerges with sharp hysteresis upon cooling, directly mirroring the resistivity loop. This observation establishes a microscopic link between the fate of the in-gap state and the macroscopic transport anomaly. Rather than being a surface effect or incidental spectral weight, the close correspondence between its hysteretic evolution and the transport hysteresis suggests that the in-gap state is intimately connected to the electronic reconstruction governing transport across the phase transition. By elevating the role of this state, our findings suggest that the C‑CDW ground state may be viewed as a partially coherent correlated phase, where strong electronic correlations coexist with a fragile but decisive conduction pathway.

The microscopic origin of the two low-energy states remains an active topic of research. The flat band observed in the C-CDW phase has traditionally been associated with the lower Hubbard band arising from strong electron correlations within the Star-of-David clusters \cite{fazekas1979electrical,pillo2000interplay,sipos2008mott}. More recent theoretical and experimental studies, however, have demonstrated that interlayer hybridization and the stacking configuration of the CDW layers can substantially modify the low-energy electronic structure, giving rise to additional states within the nominal correlation gap while preserving the correlated insulating ground state \cite{ritschel2018stacking,wang2020band,jung2022surface,park2023stacking,lee2023charge,wang2024dualistic}. Within this framework, the in-gap state observed in the present work is consistent with a reconstructed low-energy electronic structure associated with the C-CDW phase. While our ARPES measurements distinguish the flat band and the in-gap state through their different photon-energy and temperature dependences, identifying their microscopic origin unambiguously will require complementary structural characterization together with detailed first-principles calculations. The present work therefore focuses on establishing their distinct spectroscopic character and demonstrating the intimate connection between the in-gap state and the first-order C-CDW--NC-CDW transition.

In summary, our findings establish the in-gap state as a defining low-energy spectroscopic feature of the C-CDW phase. Its hysteretic collapse and recovery directly track the first-order C-CDW–NC-CDW transition and closely correlate with the corresponding transport hysteresis. These observations demonstrate that the in-gap state is an intrinsic component of the reconstructed low-energy electronic structure and underscore its important role in the electronic evolution across the phase transition. The present results provide new insight into the low-energy electronic structure of 1$T$-TaS$_2$ and establish the in-gap state as a sensitive spectroscopic probe of the C-CDW phase.

\section*{Acknowledgments}

	This research used resources ESM (21ID-I) beamline of the National Synchrotron Light Source II, a U.S. Department of Energy (DOE) Office of Science User Facility operated for the DOE Office of Science by Brookhaven National Laboratory under Contract No. DE-SC0012704. We have no conflict of interest, financial or other to declare. Author M.J. would like to acknowledge the support from NSF grant (award number 2233149) and from University of Connecticut Quantum Seed grant. Author T.Y. notes that this research also benefited from support provided by the Xiamen University Malaysia research grant (Grant No.~IPHY/0008).

\end{document}